\documentclass[twocolumn,showpacs,preprintnumbers,amsmath,amssymb]{revtex4-1}

\usepackage{graphicx}
\usepackage{dcolumn}
\usepackage{bm}

\begin{document}

\title{On the Relation between the Psychological and Thermodynamic Arrows of Time}

\author{Leonard Mlodinow}
 \thanks{Current Address: Center for Quantum Information Science and Technology, USC, Los Angeles, California.}
\affiliation{California Institute of Technology, \\
Pasadena, California, 91125
}%

\author{Todd A. Brun}
 \email{Email:  tbrun@usc.edu}
\affiliation{Center for Quantum Information Science and Technology, \\
University of Southern California, Los Angeles, California
}%

\date{\today}

\begin{abstract}
In this paper we lay out an argument that generically the psychological arrow of time should align with the thermodynamic arrow of time where that arrow is well-defined.  This argument applies to any physical system that can act as a memory, in the sense of preserving a record of the state of some other system.  This result follows from two principles:  the robustness of the thermodynamic arrow of time to small perturbations in the state, and the principle that a memory should not have to be fine-tuned to match the state of the system being recorded.  This argument applies even if the memory system itself is completely reversible and non-dissipative.  We make the argument with a paradigmatic system, then formulate it more broadly for any system that can be considered a memory.  We illustrate these principles for a few other example systems, and compare our criteria to earlier treatments of this problem.
\end{abstract}

\maketitle

\section{Introduction}

The question of why the future is distinguished from the past has been of interest ever since the nineteenth century, when physicists began to explore the fact that the equations governing dynamics are invariant under time reversal, while our environment obviously is not.  In modern physics, we say that any local, Lorentz invariant quantum field theory with positive energy is invariant under time reversal, if you also reflect in space, and interchange particles and antiparticles; together those operations generate the so-called CPT transformation. Though we now know that C and P can sometimes be violated, in most practical applications, those violations play no role, and so the dynamics will be invariant under T.  

Despite the interchangeability of past and future with respect to the laws of microscopic physics, we humans have no trouble distinguishing whether time is moving forward or backward.  Smoke rises and disperses from chimneys, but never gathers and returns; dropped eggs splatter on the floor, but once dropped they never jump back into their shell.  This time asymmetry can be understood as a result of the second law of thermodynamics: it follows from the equations of physics governing time evolution that if a system is confined to a small region of phase space at some time, then with virtual (but not total) certainty it will occupy a much larger region of phase space at other times.  The observed asymmetry of past and future can therefore arise due to a boundary condition that, in one direction of time, near the big bang, the universe was in a state of low entropy.   Given that situation, we can define the ``past'' of any given moment as those times that are closer to the big bang, when the entropy of the universe was lower, and the future as those times that are further from the big bang, and in which the entropy of the universe is higher.  This is the ``thermodynamic arrow of time.''

Beside watching whether eggs are splattering into messes or cleaning themselves up, there is another way we can define a direction in time: we remember the past, but not the future.  From the point of view of human psychology, this is perhaps the key to our feeling for past and future.  After all, though it would cause raised eyebrows if we observed smoke retreating into chimneys, most of us would be far more surprised if a memory of that chimney turned out to reflect the chimney's future state rather than its past.  Because of this, physicists sometimes refer to the direction of our memory as the ``psychological arrow of time'' (see, for instance, Hawking, 1985 and 1994) \cite{Hawking85,Hawking94}.

The thermodynamic and psychological arrows of time obviously agree, but why should they?  If the equations that govern a system are agnostic with regard to which way is the future, then why do systems (memories) arise in nature that reflect the state of other systems (the systems being remembered) in one direction of time, but not the other?  Why do streaks in mica correlate with cosmic rays that traversed them in the past (as defined by the thermodynamic arrow) but not the future?  Why do we see contrails from planes past, but not planes yet to come?  Why do we remember the thermodynamic past, but not the thermodynamic future? 

Imagine that our universe were in a state such that, at some time $T$ in the future (i.e., further in time from the big bang than we are today) the entropy were to start to decrease.  It is extremely unlikely that we would be in such a state, but it does not violate the fundamental laws of physics.  Then for times greater than $T$, entropy would decrease with time rather than increasing.  For $t>T$ would people the world over marvel at chimneys sucking back plumes of smoke, and broken eggs jumping into their shells, or would the psychological arrow of time also reverse itself, with the result that we perceived time to run in the reverse direction, and the second law still held true?  Our answer is that latter. In what follows, we will show that the psychological and thermodynamic arrows must point in the same direction: the psychological arrow follows from the thermodynamic one.

That the psychological and thermodynamic arrows must align has in the past been argued on the basis of Landauer's principle \cite{Landauer61}, which states that the total entropy of a closed system must increase upon the erasure or reinitialization of a memory record.  According to that argument \cite{Hawking85,Hawking94,Wolpert92,Hartle05,Hartle13} a realistic memory must allow for the erasure of records.  Erasure, Landauer showed, is in essence like the smashing an egg---it is a process whose time reversal violates the second law.  A memory, in that view, must therefore be a dissipative, hence irreversible, system, and so to remember the future would be a feat akin to orchestrating the remnants of a smashed egg to jump back and reassemble inside its shell.  We will argue, on the other hand, that the principle that the psychological and thermodynamic arrows of time must align actually arises from a combination of the fundamental microscopic laws of physics and a reasonable requirement of what it means for a system to function as a memory.  Therefore, the two arrows must align even for systems that are time reversible.

In Section II we present a simple paradigm for a memory embedded in a system with a well-defined thermodynamic arrow of time.  The dynamics of this system---including the memory---are reversible, but we argue that the direction of memory must still match the thermodynamic arrow of time.  In Section III we give a more precise definition of what it means for a system to act as a memory, and give a broader argument that the direction of memory must align with the arrow of time.  In Section IV we examine a few other examples of systems that can function as memories, and argue for generic differences between memory (i.e., correlations with the past) and anticipation or projection (i.e., correlations with the future).  In Section V we compare our argument to other takes on this problem, and discuss their points of commonality, and in Section VI we conclude.


\begin{figure}[htbp]
\begin{center}%
\includegraphics[width=3in]{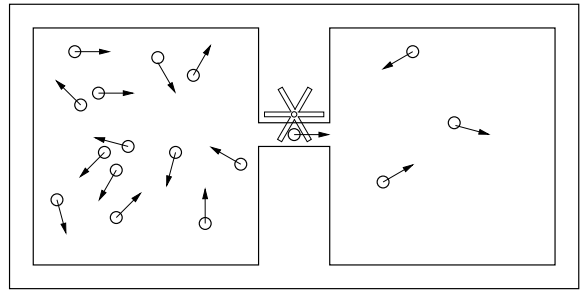}
\caption{A reversible system with a well-defined arrow of time and a memory.}
\label{vessel}
\end{center}
\end{figure}


\section{A simple paradigm}

Consider the system depicted in figure~\ref{vessel}.  A sealed vessel is divided into two chambers, separated by a barrier with a single narrow gap.  Within the vessel are $N$ elastic particles that can collide with each other or the walls, but are otherwise free.  In the gap between chambers is a rotor---a reversible counter, or turnstile---that rotates one position clockwise or counterclockwise as a particle moves left or right, respectively.  This system should be considered closed---it is not in contact with a thermal bath---and all the dynamics are reversible.  Nevertheless, generically this system can exhibit a thermodynamic arrow of time if at some early time, $t_{\rm i}$, there is an imbalance of particles between the two chambers, as is depicted in the figure. From almost all states with such an imbalance, the system will tend to evolve toward a state in which the number of particles in the two chambers are roughly equal.  A thermodynamic arrow of time is defined for systems in which coarse-grained variables---such as, in this case, the total number of particles on each side---exhibit a preferential direction of evolution.  For times greater than $t_{\rm i}$, therefore, the thermodynamic arrow of time of this system will point toward increasing times, which we will call the future.  This is like the present situation of our universe: we are in a nonequilibrium state with a boundary condition that at some time the universe was in a state of much lower entropy; we call that direction of time the past.       

The rotor in this system can function as a memory that records the net number of particles that crossed from the left chamber to the right.  Suppose that the rotor has $M$ positions labelled $0,\ldots,M-1$, and that (for simplicity) the number of distinct positions $M$ is much greater than the average number of particles that will cross between $t=t_{\rm i}$ and a later time $t=t_{\rm f}$.  We let $r(t)$ be a coarse-grained variable that takes the values $0,\ldots,M-1$ corresponding to the rotor position at time $t$.  We also assume that the number of particles is sufficiently small (compared to the size of the vessel) that the probability of more than one particle crossing at the same time is negligibly small, and that the average energy of the particles is sufficiently low that the possibility of fast particles that might transfer enough momentum to make the rotor spin (rather than just shifting one position) can also be neglected.

We `read out' the record of the net passing of particles between time $t_{\rm i}$ and a time $t$ by using a simple function
\begin{equation}
f_{\rm read}(r(t)) = r(t) - r_{\rm ref}\mod M ,
\label{rotor_past}
\end{equation}
where $r_{\rm ref}$ is the setting of the rotor at the reference time, $r_{\rm ref} = r(t_{\rm i})$.  This read-out function provides a memory of the past.  

Because this whole system is deterministic and reversible, it seems as if we could equally well choose a reference time $t_{\rm f}>t$ and interpret the rotor as a memory {\it of the future} by defining a different `read-out' function:
\begin{equation}
f'_{\rm read}(r(t)) = r'_{\rm ref} - r(t) \mod M ,
\label{rotor_future}
\end{equation}
where $r'_{\rm ref} = r(t_{\rm f})$. At times $t<t_{\rm f}$, we can now interpret the rotor as a memory of the net number of particles that {\it will} cross from left to right between now and the later time $t_{\rm f}$.  Our simple memory seems to have become a memory of the future.    

One might object to describing the rotor as a memory of the future on the grounds that since $t_{\rm f}>t$ we would not generally know, at time $t$, the value of $r'_{\rm ref}$, and hence could not use the rotor as described in (\ref{rotor_future}).  That argument depends on the implicit assumption that ``time flows,'' which immediately implies that we cannot remember the future because ``it hasn't happened yet.''  That naive view is unsatisfactory, because the microscopic dynamics are reversible and give no privileged role to either direction in time.  There is actually no a priori reason that a rotor coupled as we've described should not reflect the state of the particles at future times, just as it does for the past.  In fact, any system with reversible dynamics encodes both past and future within in it, as Laplace \cite{Laplace} pointed out over two centuries ago:

\begin{quote}
We may regard the present state of the universe as the effect of its past and the cause of its future. An intellect which at a certain moment would know all forces that set nature in motion, and all positions of all items of which nature is composed, if this intellect were also vast enough to submit these data to analysis, it would embrace in a single formula the movements of the greatest bodies of the universe and those of the tiniest atom; for such an intellect nothing would be uncertain and the future just like the past would be present before its eyes.
\end{quote}

Our key observation is that, despite the fact that the state of a classical system encodes both its past and future, when the system being remembered has a well-defined thermodynamic arrow of time, we can rule out the possibility of a memory recording its future, because any such memory {\it could remember only one possible configuration of that system}.  If the system changed its state even minutely, the memory would no longer create an accurate record.  We call the requirement that a memory be capable of remembering more than one fixed state of a system {\it generality}.

Imagine you have a digital camera that contains a chip capable of recording what the camera's sensors pick up, but only for one particular scene.   Suppose, further, that every time you change any aspect of the scene, you have to insert a new chip designed specifically to record that precise scene.  Heuristically, one would think that a chip of that sort could hardly be called a memory, for a memory that must be fine-tuned based on the state of the system being remembered is hardly of any use.  But there is another, more fundamental reason to require generality: it allows us to rule out, as memories, systems that are {\it correlated} with other systems but are not causally related.   

Consider the system we described above.  We can, in theory, calculate the state of the particles at all times, and hence, in place of the rotor, we could have set up an independent meter that is not coupled to the particles at all, but which we have designed (based on our calculation) to always exhibit the correct number of net particles crossing between the chambers as determined by our calculation.  A meter such as that, which we set up to reflect the particle crossings, but which does not interact with the particles in any way, should not be considered a memory, yet would have the same correlations with the system of particles as the rotor.  The key difference between the two is that a small alteration in the state of the particle system would not destroy their correlation with the rotor, whereas it would destroy the correlation with the independent meter we set up.  Unless we redid our calculations and made a compensating alteration in the meter, it would no longer accurately reflect the passage of the particles.  Correlated systems such as the meter are excluded from the definition of memories by the principle of generality:  a memory should be capable of remembering more than one thing.

If we apply this idea to our rotor, we immediately see that there is a sharp distinction between the two functions $f_{\rm read}(r(t))$ and $f'_{\rm read}(r(t))$, in (\ref{rotor_past}) and (\ref{rotor_future}), that describe the rotor's read-out.  

To see this, let us begin by supposing, in the case of $f_{\rm read}(r(t))$, that we slightly alter the state of the particles (but not the rotor) at $t_{\rm i}$.  We alter the coordinates and momenta of the particles by a small amount, but such that no particle changes position from one side of the barrier to the other, or, more precisely, so that the coarse-grained variable $r(t_{\rm i})$ remains unchanged.  If we then let the system propagate from $t_{\rm i}$ until $t$ under the action of the same Hamiltonian as in the unperturbed case, the system will in general exhibit the same arrow of time, and though its time evolution will be altered, the read-out function in (\ref{rotor_past}) will remain an accurate reflection of the past.  In this case, therefore, the rotor functions as a memory of the past for a large family of states that are in some sense ``near'' the unperturbed state we discussed above.        

Now let's consider the other case.  In this case, when we look for a family of ``nearby'' states for which the rotor continues to function as a memory of the future, we find that, generically, we must reject them all because they do not preserve the arrow of time---that is, even the most minute generic changes in the coordinates of the system at $t_{\rm f}$, when propagated back to earlier times will result in a system with a more equal distribution of particles than at $t_{\rm f}$, rather than a more lopsided distribution. 

The unperturbed system, due to our assumption of the existence of the thermodynamic arrow of time, evolves forward in time from a more lopsided distribution of particles at $t_{\rm i}$ to a more evenly distributed system at $t_{\rm f}$; the time reverse of that evolution will therefore propagate the system from a more even distribution to a less even one.  But that state of the unperturbed system at $t_{\rm f}$ was, due to the way it arose, a very special state, and one that is unstable with respect to perturbation.  A system of particles in a {\it generic} state at a time $t$ will lead to a state with a more even distribution of particles at {\it both} later and earlier times.  Hence, when it comes to remembering the future, there is no family of ``nearby'' states that is both consistent with the existence of an arrow of time, and for which the rotor continues to function as a memory of the future.  As a memory of the future, the rotor does not satisfy generality.  

These principles can be readily generalized beyond this simple system, and we do so in the following section.  The conclusion is clear:  generically, we expect the `psychological arrow of time' (broadly defined for any system that can act as a `memory') to align with the thermodynamic arrow of time in any system where that thermodynamic arrow is well-defined.

\section{What is a memory?}

It is difficult to come up with a simple criterion that is general enough to capture every example of a system that can function as a memory.  However, we will propose a definition that seems to apply to many examples, and that satisfies the criterion of ``generality'' discussed in the previous section.

Suppose we can divide the world into two subsystems:  the {\it record} (or memory) $R$, whose state is given by a vector $\mathbf{r}$ in some phase space $\mathcal{R}$, and the {\it system}  $S$ (effectively, whatever is being recorded plus everything else) whose state is given by a vector $\mathbf{s}$ in some phase space $\mathcal{S}$.  We will assume for the moment that the system and record are classical and completely reversible.  The system and record evolve together.  We can determine (in principle) the values of $\mathbf{s}(t)$ and $\mathbf{r}(t)$ for all times $t$ by specifying the values $(\mathbf{s}_0,\mathbf{r}_0)$ at any one time $T$.  At the moment, we make no assumptions about whether $T$ is in the past of future of any of the other relevant times (e.g., $t_{read}$, $t_1$, and $t_2$ in the conditions below).

For the subsystem $R$ to serve as a memory, it must satisfy the following requirements:

\begin{enumerate}
\item We can define two vectorial functions $\mathbf{f}_R(\mathbf{r}(t))$ and $\mathbf{f}_S(\{\mathbf{s}(t)\},t\in I)$, where $\mathbf{f}_R$ is a function of $\mathbf{r}$ at some time, and $\mathbf{f}_S$ is a function of $\mathbf{s}$ over some interval or range of times $I$.

\item There is some read-out time $t_{read}$ and interval $I=[t_1,t_2]$ such that $\mathbf{f}_R(\mathbf{r}(t_{read})) \approx \mathbf{f}_S(\{\mathbf{s}(t)\},t\in I)$.  That is, at the time $t_{read}$ the state of the record reflects the state of the system at some time, or range of times. 

\item {\it Generality}.  Suppose that we change the state of the system $\mathbf{s}_0$ at $T$ while keeping the state of the record subsystem $\mathbf{r}_0$ fixed.  We require that there be some nontrivial set of possible $\mathbf{s}_0$s such that condition 2 above remains true, without changing the definitions of $\mathbf{f}_R$ or $\mathbf{f}_S$, and without changing the time $t_{read}$ or the interval $I$.  We also require that the functions $\mathbf{f}_R(\mathbf{s}(t_r))$ and $\mathbf{f}_S(\{\mathbf{s}(t)\},t\in I)$ are not constant over this set of possible $\mathbf{s}_0$s.

\item {\it Thermodynamic robustness}.  The read out and coarse graining functions $\mathbf{f}_R(\mathbf{r}(t))$ and $\mathbf{f}_S(\{\mathbf{s}(t)\},t\in I)$ must be robust under small perturbations of the microscopic degrees of freedom (of both the system and the memory) at $T$.
\end{enumerate}

Let us examine these conditions one at a time.  In condition 1, the function $\mathbf{f}_R$ is a read-out function.  It specifies how the record is encoded in the microscopic degrees of freedom of the subsystem $R$.  Similarly, $\mathbf{f}_S$ extracts whatever property of the system $S$ is being stored.  In general, these functions will involve a great deal of coarse-graining over the fine details of their respective subsystems.  

Condition 2 gives the obvious requirement that the read-out of the record should indeed correspond to whatever quantity is supposed to be recorded.  In the condition here, we have required only that this correspondence be approximate.  Another reasonable way of defining this would be to require that the two functions correspond for many states $\mathbf{s}_0$, but not necessarily all.  Note also that we have treated a memory that is read at a single moment of time.  It is straightforward, however, to extend this definition to memories that operate over a range of times, and even to memories whose records evolve as they interact with the system.  Here we consider only the simplest case.  

We can see how the rotor system in the previous section satisfies conditions 1 and 2.  The function $\mathbf{f}_R(\mathbf{r})$ is the read-out function $f_{\rm read}(r)$.  We can define the corresponding variable $\mathbf{f}_S$ straightforwardly.  Let $N(\mathbf{s})$ be the number of particles to the right of the partition when the underlying variables are $\mathbf{s}$.  (This is a simple example of a coarse-graining.)  Then $\mathbf{f}_S(\{\mathbf{s}(t)\},I)$ is simply $N(\mathbf{s}(t)) - N(\mathbf{s}(t_{i}))$ in the case of remembering the past, and $N(\mathbf{s}(t_{f})) - N(\mathbf{s}(t))$ in the case of remembering the future.   

Condition 3 embodies the assumption of {\it generality}:  In order to consider $R$ a memory, it must be capable of recording different values for different states.  The requirement that $\mathbf{f}_R$ and $\mathbf{f}_S$ are not constant guarantees in addition that the correspondence between $\mathbf{f}_R$ and $\mathbf{f}_S$ is nontrivial---that is, we exclude the possibility that $\mathbf{f}_R$ is effectively fixed, and $\mathbf{s}_0$ is restricted to a set with some constant property.  Condition 3 thus requires that the correlation between $S$ and $R$ is not due to fine-tuning.  That is, the system and memory must actually interact, and the memory must encode some information about the system, rather than both being correlated with some common cause.

When we applied condition 3 to the rotor system, we chose $T$ to be the reference time, $t_i$ (in the case of remembering the past).  That is because if $T$ were chosen to be any other time, generic perturbations to $S$ at $T$ would alter the number of particles on either side of the box at $t_i$, and hence the function $f_{\rm read}(r)$ would have become invalid.  As a result, in that example, in order for the same function $f_{\rm read}(r)$ to work for different $\mathbf{s}_0$s, we chose the reference time $T$ to match the reference time in Eq.~(\ref{rotor_past}).  

Condition 4 is a reflection of our assumption that there exist a thermodynamic arrow of time.  The question of whether a memory can record the future makes no sense otherwise, since it the the thermodynamic arrow that we use to define past and future.  Condition 4 guarantees this.  Consider the first case of the rotor system and memory, for which Condition 3 required that $t=T=t_{\rm i}$.  Suppose that, at that time, the system is in a configuration like that pictured, with far more particles on the left side of the partition.  From almost all initial conditions with such an imbalance, the system will tend to evolve toward a state in which the number of particles in the two chambers are roughly equal.  For times greater than $t_{\rm i}$ the thermodynamic arrow of time of this system therefore points toward increasing times, which we have called the future.  This is like the present situation of our universe: we are in a nonequilibrium state, with a boundary condition that at some time the universe was in a state of much lower entropy; we call that direction of time the past.  

Though perturbing the system we just described will generally not destroy the thermodynamic arrow of time, there are very special conditions for which such a perturbation would have that effect.  For example, consider a system which, as time increases from $t_{\rm i}$, evolves into a state where the numbers of particles are more unequal.  This is not a highly intuitive choice, but in fact we can find such conditions (in principle) by following the usual intuitive evolution, and then ``running the film backwards.'' Because the dynamics are reversible, this is a legal (albeit unlikely) evolution.  We denote this time-reversed initial condition $\bar{\mathbf{s}}_T$, where we use the bar to denote the time-reversed version of the phase-space vector $\mathbf{s}(T)$.

Initial conditions $\bar{\mathbf{s}}_T$ are analogous to states in which shattered tea cups fly back together.  Starting from $\bar{\mathbf{s}}_T$, the distribution of particles between the two chambers becomes increasingly more unequal, rather than less.  This system exhibits an arrow of time, but the arrow for such a system is not robust: almost all states except for a tiny neighborhood of $\bar{\mathbf{s}}_T$ will {\it not} exhibit this behavior; and the longer we want this unintuitive behavior to persist, the smaller this neighborhood must be.  Even an initial condition $\bar{\mathbf{s}}_T'$ that is very close to $\bar{\mathbf{s}}_T$ will tend to exhibit the generic behavior, in which the distribution of particles becomes more rather than less equal.

For the rotor system, condition 3 thus requires that the reference time $T$ be chosen to coincide with $t_{\rm i}$ or $t_{\rm f}$, while condition 4 eliminates the latter.  $T$ must be an ``initial condition'' in the sense of the thermodynamic arrow of time.  So if $t_{\rm i}$ is a time exhibiting a strong deviation from equilibrium (leading to a well-defined thermodynamic arrow of time in the direction of increasing $t$), then $T$, the interval $I=[t_1,t_2]$ over which the system is recorded, and the read-out time $t_{read}$, must satisfy
\[
t_{\rm i} \le T < t_1 < t_2 < t_{read} .
\]
In other words, even if the memory subsystem is itself a completely reversible system, its ``psychological'' arrow of time must align with the thermodynamic arrow of time.  If it does not, it will not satisfy the generality or robustness requirements that we have argued must be part of the definition of a memory.

\section{Examples of systems that function as memories}

Having introduced the concepts underlying our definitions with the model system in Sec.~II, we can now try to apply these ideas to other examples of systems that function as memories.  While in most cases we cannot exactly specify the readout and coarse-graining functions $\mathbf{f}_R(\mathbf{r}(t))$ and $\mathbf{f}_S(\{\mathbf{s}(t)\},t\in I)$, it is not difficult to argue for the existence of these functions, nor is it difficult to argue that they satisfy the requirements specified in section III.

Most obvious examples of memories are synthetic---that is, they are systems deliberately engineered by humans to store information of a particular type.  We will briefly consider some canonical examples of this below.  But certain naturally-occurring systems also satisfy the definition of a memory given above.  Most of these examples depend on irreversible processes, so it is not surprising that they align with the usual thermodynamic arrow of time \cite{Hawking85,Hawking94,Wolpert92,Hartle05}.  But in principle this is not necessary---reversible systems can also function as memories, and when they do so, their arrows of time will also align with the thermodynamic arrow, by the argument in section III.

\paragraph{Synthetic memories:  computer storage.}  Given the ubiquity of modern electronic technology, various kinds of computer memories are all around us.  These include computer RAM; magnetic storage on hard drives (and, still in a few applications, magnetic tape); flash memory; and charge-coupled devices (CCDs) and active-pixel sensors (APSs) in digital cameras.

If we take as an example a single computer bit, the readout function $\mathbf{f}_R(\mathbf{r}(t))$ corresponds to a range of average voltages across the bit.  A ``high'' range might be on the order of 5 V, the ``low'' range close to 0 V.  The bit stores an input voltage that was applied at an earlier time, so $\mathbf{f}_S(\{\mathbf{s}(t)\},t\in I)$ in this case is also an averaged voltage range. It is clear that these functions satisfy the requirements in section III:  the computer bit can store either 0 or 1, depending on its input, and no fine tuning between the input and the bit is necessary.  Moreover, the value of the stored bit is correlated with the input at an earlier time, not with later inputs, as common sense would suggest.

\paragraph{Reversible and quantum computers.}  Existing computer technology is emphatically not reversible:  anyone who has held a laptop computer actually on his or her lap is quickly aware of how much heat such a computer produces.  But it was argued by Landauer \cite{Landauer61} and proven by Bennett \cite{Bennett88} that dissipation is not necessary, in principle, for computation.

Bennett's proof involved first defining a reversible model of a Turing machine.  A Turing machine is an abstract model of a computer, and has two subsystems that could be considered memories by the above definition.  First, the computer has a {\it tape} (or more generally $n$ tapes), divided into unit cells, each cell containing one symbol from a finite alphabet.  The ``read/write head'' of the Turing machine is located at a particular cell at any given time.  Second, the computer has an {\it internal state}, which is one of a finite set of possible settings.  The Turing machine is designed with a transition rule, or program:  given the value at the current tape cell and the current internal state, the machine replaces the symbol with a new symbol, moves the head at most one position to the left or right, and transitions to a new state.

In Bennett's reversible model, the transition rule is invertible:  given the current tape symbol, location and internal state it is possible to invert the transition rule and return to the state of the Turing machine at the previous state.  He proved that such a reversible machine could emulate a standard Turing machine, at the cost of a relatively small overhead in the amount of tape used.

From the work of Landauer, such a logically reversible machine can also be made to be physically reversible.  Moreover, in principal such a machine need not dissipate any power in order to function.

Of course, in practice even such a reversible memory generally would be used by first setting it in a standard starting state.  For instance, a string of $n$ reversible bits would be prepared in the $00\cdots0$ state.  They could then reversibly store $n$ input bits $x_1 x_2 \cdots x_n$ by being XORed with those bits; this procedure would simply copy the values $x_1 x_2 \cdots x_n$ into the memory.  This initialization step would be an irreversible process.

But in principle this initialization step to $00\cdots0$ is not necessary.  Suppose that the reversible bits started instead in a general state $y_1 y_2 \cdots y_n$.  Then, after being XORed with the input bits $x_1 x_2 \cdots x_n$, the memory would be left in the state $(x_1\oplus y_1)(x_2\oplus y_2)\cdots(x_n\oplus y_n)$.  However, it is still quite possible to recover the stored bit values simply by redefining the readout function to take the initial state of the reversible bits into account.  The readout function would be $\mathbf{f}_R(z_1 z_2 \cdots z_n) = (z_1\oplus y_1)(z_2\oplus y_2)\cdots(z_n\oplus y_n)$, where $z_1 z_2\cdots z_n$ is the state of the reversible bits at the time of readout.  We can see that this readout function satisfies the requirements of section III:  changing the system being recorded (in this case, the input bits $x_1 x_2 \cdots x_n$) while keeping the initial state of the memory fixed does not require us to change the readout function $\mathbf{f}_R$.

Reversible computation may be of practical interest for classical computers to reduce the problem of power dissipation; but it is absolutely vital for quantum computers, whose unitary operation must be intrinsically reversible.  While our definition in section III probably cannot be applied to quantum systems without alteration (more on this below), we believe the general principles requiring that memories align with the thermodynamic arrow of time will apply to quantum memories as well.

\paragraph{Photographic film.}  Photographic film is a canonical example of a synthetic, irreversible system designed (unlike computer memories) to produce a permanent record.  To produce a photograph requires that the film be prepared in a very special initial state.  This preparation step will clearly be irreversible.  The record is produced by exposing the film to light, thus driving irreversible chemical reactions.  Since both the preparation step and the recording step involve irreversible processes, the ``arrow of time'' associated with this record must clearly line up with the thermodynamic arrow of time.  This type of irreversible record is used as a paradigm in the work of Wolpert \cite{Wolpert92}.

\paragraph{Tracks in mica and contrails in the stratosphere.}  The examples considered so far are all synthetic systems, designed to record and retrieve information.  Of course, naturally occurring systems can also function as memories.  Indeed, any two systems that interact have to potential to exchange information about their respective states.

In most cases, however, this information is spread out in the form of complicated correlations between many degrees of freedom in a way that makes it impractical to retrieve.  Gell-Mann and Hartle refer to such correlations---perhaps retrievable in principle, but certainly not in practice---as {\it generalized records} \cite{Gell-MannHartle98}, to contrast them with the type of synthetic records described above.

There are at least a few cases, however, where naturally occurring systems can store information for some length of time in a form that is robust and not particularly difficult to retrieve.  A classic example is the existence of cosmic ray and fission product tracks in naturally occurring crystals of mica \cite{PriceWalker62}.

The passage of high-energy particles through regular crystals of mica disrupts the crystalline structure along the trajectory of the particle.  These tracks can be made visible by etching the crystal in acid.  (In fact, synthetic crystals of mica have been used as simple particle detectors.)  In this case, both the ``preparation of the memory''---that is, the formation of the crystal---and the recording process are irreversible.  In this case, however, the preparation step occurs naturally (attesting to the nonequilibrium state of the earth's crust).  The read-out function---the tracks of disruption---record the number of high-energy particles and their directions, and perhaps some information about their energy, but not in general when they occurred.

\paragraph{Ripples in water and other outgoing waves.}  For completeness, it would be good to consider a naturally occurring ``memory'' system that was also reversible.  Most such records are unfortunately transient in their existence.  However, a familiar phenomenon does illustrate this type of system---namely, the emission of waves from moving (or reflective) objects.

The canonical example of this is the spreading of ripples from a stone tossed into a pond.  These ripples are not perfectly reversible, of course, but dissipation is not essential to their function.  Nor, in principle, does the pond have to be prepared in an exact state---waves are generically linear, and hence could be separated from a background.

Our own senses make use of sound and light waves to acquire information about objects outside ourselves.  Given their finite speed of propagation, any wave we intercept is in reality a record of an earlier event.  While in everyday circumstances, these records are not of long standing, we are quite capable of seeing light emitted from stars thousands of years ago---or from other galaxies millions or (with the help of telescopes) even billions of years ago.

The propagation of waves is generally reversible to a good or even excellent approximation.  In principle, the emission process is also reversible.  Nevertheless, wave emission almost always exhibits a strong arrow of time.  This is because the universe is far from equilibrium, which in turn is due to the overall thermodynamic arrow of time arising from the fact the the universe apparently began in a very low entropy state.

\section{Discussion}

\subsection{Memory vs. Anticipation}

As we have pointed out above, it is quite possible for a system at a given time to be correlated with the state of another system in their thermodynamic future rather than their past.  We argue that such correlations are inconsistent with the properties we expect of a ``memory,'' as defined in Section III above.  However, even in a system with a well-defined thermodynamic arrow of time, such correlations can and do arise; we would term such a correlation with the future state of another system ``anticipation'' or ``prediction.''

Of course, for systems with very regular dynamics the future is highly predictable---these systems exemplify Laplace's idea that the present state reveals the future as well as the past.  But even in less trivial systems, it is possible to anticipate future behavior up to a point.  We expected such correlations to differ from memories, however, in a number of respects.

Consider our paradigmatic example from Section II.  By counting the net number of particle transitions from the left half to the right half of the vessel, one can estimate a rough transition rate between the sides.  This, in turn, would allow one to extrapolate the number of particles on each side in the future.  If the numbers of particles on each side at the reference time $T$ are $N_L(T)$ and $N_R(T)$, respectively, then we would project the numbers at some future time $t>T$ to roughly follow an exponential law:
\begin{eqnarray*}
N_L(t) &=& \frac{1}{2} \left[ N_L(T) \left(1+e^{-2\gamma t}\right) + N_R(T)\left(1-e^{-2\gamma t}\right)\right] ,\\
N_R(t) &=& \frac{1}{2} \left[ N_L(T) \left(1-e^{-2\gamma t}\right) + N_R(T)\left(1+e^{-2\gamma t}\right)\right] .
\end{eqnarray*}
The transition rate $\gamma$ can be estimated from observation.  If we let the system evolve for some length of time $\Delta t$ and observe $n$ net particle transitions, we would estimate $\gamma\approx n/\Delta t(N_L(T)-N_R(T))$.   We expect, however, that this estimated rate will not be precisely accurate, and of course even if it were, the actual behavior will deviate from the expectation.  Thus, we expect the accuracy of this future correlation to fall off exponentially, while the accuracy of the past correlation remains precise.  (Of course, at very long times the system will approach equilibrium, with a roughly equal distribution of particles between the two chambers.)

We can compare this difference to our experience of the real world as well.  Information that is stored in a robust record can be retrieved very well, even after a long period of time; and how well it can be retrieved is not dependent on the regularity of the system.  For example, we known many details of the last day of Julius Caesar's life, more than 2000 years later.  But we are unable to predict in any detail the days of the most famous of our current citizens even a short time into the future.

\subsection{When is ``Now?''}

Our psychological perception of time as ``flowing'' implies that there is a special time---the present, or ``now''---at which events in the future undergo a profound transition and become events in the past.  This feeling is so fundamental that it is difficult to conceive of any other way that time could be experienced.  And yet, the physical description of systems evolving in time does not include this notion of ``now.''  All times are treated with the same status; none is singled out; and no physical principle seems to imply that time must be experienced in this way.

The modern interpretation of this dividing line between the past and the future is that it is psychological in origin.  At all moments, we have memories of the past, but not of the future, which makes a profound difference to how we regard past and future.  As argued in the previous section, records of the past can remain robust and reliable for long periods with little loss, while our ability to anticipate the future falls of extremely rapidly.  Psychologically we make a very sharp distinction between these two processes, which is no doubt an adaptive trait.

However, the notion of ``now'' as an idealized point between the past and the future does not really hold up to scrutiny \cite{Hartle05}.  The process of registering and recording a memory will in general take some characteristic time, so that our conscious impression of when something happens will in general lag the actual event by some amount.  Our brain does its best to stitch all its sense impressions together into a seamless whole, so we are in general not aware of any lag.

Moreover, we have a very good ability to anticipate highly regular events in the immediate future.  Anyone who has ever caught a ball or stepped onto a moving escalator has experienced this:  to catch, we reach not for where the ball is, but where it will be.  We can therefore think of the moment ``now'' as actually being somewhat spread out in time, with some lag into the (idealized) past and possibly a small extension into the (idealized) future.

\subsection{Conclusions}

One of our deepest observations about nature is that we remember the past, but not the future:  that is, that the psychological arrow of time aligns with the thermodynamic arrow of time.  In fact, this phenomenon is so deeply embedded in our experience that it took almost all of history even to recognize that there was a question to be answered.

Our modern understanding of the arrow of time recognizes its origins in an unusual, low entropy state of the universe in the far past.  This leads to the thermodynamic arrow of time, exhibited by irreversible systems, which includes almost all systems of sufficient size and complexity.  Since almost all physical systems that can function as memories or records are themselves irreversible---either in their dynamics, or in the requirement of an irreversible preparation step, or both---it is perhaps not surprising that they exhibit an arrow of time that aligns with the overall thermodynamic arrow.

In this paper, however, we have argued that even completely reversible systems that can function as memories must exhibit an arrow of time that aligns with the usual thermodynamic one.  This arises purely from a reasonable notion of what it means for a system to be a memory:  that it must exhibit correlations with another system; that these correlations must arise not from careful fine-tuning of the system and memory together, but from interactions between them; and that these correlations must be robust to small perturbations in the system and memory states.  A memory is only a memory if it has the potential to remember more than one thing.  We illustrated this argument with a simple paradigmatic system, and then pointed out how it works in a variety of examples.

There is a very important open question left in this work.  The models we have chiefly focused on, and our definition of a memory, take an implicitly classical view of the world:  we assume that the state of the universe can be factorized, so that the system and memory can each be treated as having a definite well-defined state at any given time.  But of course, the universe is actually quantum mechanical.  This definition cannot be applied directly to quantum systems, because of the phenomenon of entanglement, which implies that the system and memory states cannot necessarily be factored at all times.  We believe that the essence of this argument will still hold for quantum systems as well as classical ones; but the technical details of how to make this argument precise are work for the future.

In the meanwhile, we believe that our current work validates our intuition about the arrow of time, and gives insight into what it means to remember the past, and to dream about the future.

\section*{Acknowledgments}

LM and TAB would like to acknowledge fascinating and useful conversations over the years with Sean Carroll, Murray Gell-Mann, Jonathan Halliwell, Jim Hartle, Stephen Hawking, and John Preskill.  This work was supported by the hospitality of the Institute for Quantum Information at Caltech.

\end{document}